\documentclass[a4paper,11pt]{article}
\pdfoutput=1
\usepackage{jcappub}

\usepackage{color}
\usepackage{amsfonts}
\usepackage{graphicx}
\usepackage{amssymb}
\usepackage{amsmath}

\usepackage{eufrak}
\usepackage{mathrsfs}

\renewcommand{\thesection}{\arabic{section}}

\def\laq{~\raise 0.4ex\hbox{$<$}\kern -0.8em\lower 0.62ex\hbox{$\sim$}~}
\def\gaq{~\raise 0.4ex\hbox{$>$}\kern -0.7em\lower 0.62ex\hbox{$\sim$}~}

\def\beq{\begin{equation}}
\def\eeq{\end{equation}}
\def\bea{\begin{eqnarray}}
\def\eea{\end{eqnarray}}

\def \ra {\rightarrow}

\def \Da {\Delta}
\def \b {\beta}
\def \a {\alpha}
\def \ap {\alpha^{\prime}}

\def \sg {\sigma}
\def \da {\delta}
\def \ep {\epsilon}
\def \r {\rho}

\def \Mp {M_{\rm P}}
\def \Ms {M_{\rm S}}

\title{On the initial regime of pre-big bang cosmology}

\author{M. Gasperini}

\affiliation{Dipartimento di Fisica, Universit\`a di Bari, \\ 
Via G. Amendola 173, 70126 Bari, Italy\\
and Istituto Nazionale di Fisica Nucleare, Sezione di Bari,\\ Via E. Orabona 4, 70125 Bari, Italy}

\emailAdd{gasperini@ba.infn.it}

\abstract{The production of a background of super-horizon curvature perturbations with the appropriate (red) spectrum needed to trigger the cosmic anisotropies observed on large scales is associated, in the context of pre-big bang inflation, with a phase of growing string coupling. The extension towards the past of such a phase is not limited in time by the dynamical backreaction of the quantum perturbations of the cosmological geometry and of its sources. A viable, slightly red spectrum of scalar perturbations can thus be the output of an asymptotic, perturbative regime which is well compatible with an initial string-vacuum state satisfying the postulate of ``Asymptotic Past Triviality".}

\keywords{cosmological perturbation theory
 
\vskip13pt plus8pt minus11pt

\noindent{\bfseries\large\sffamily{Preprints:}} BA-TH/712-17
}

\begin{document}

\maketitle


\section{Introduction}
\label{Sec1}
\setcounter{equation}{0}

In the context of pre-big bang cosmology \cite{1,2}, the quantum fluctuations of the metric tensor are generically amplified with a ``blue" spectrum \cite{3,4} which is too steep to give any currently appreciable contribution at the large-distance scales relevant to the present CMB observations. In that context, the adiabatic scalar perturbations responsible for the observed CMB anisotropy \cite{15} are produced -- via the so-called ``curvaton'' mechanism \cite{5,6,7} -- by the amplification of the Kalb-Ramond axion field $\sg$ \cite{8,9}, associated by space-time duality with the four-dimensional components of the NS-NS two-form present in the bosonic sector of the (dimensionally reduced) string effective action.

In fact, thanks to the minimal coupling of the axion fluctuations to the scalar metric perturbations, the dominance and the subsequent decay of the axion $\sg$ automatically induces a super-horizon spectrum of adiabatic curvature perturbations, $\Da_R^2(k)$, whose slope is exactly the same as the one of the primordial axion spectrum, $\Da_\sg^2(k)$. Such a slope, depending on the details of the pre-big bang dynamics, can be flat enough \cite{10,11} -- and even slightly red \cite{12,12a,13,14} -- just to be in full agreement with the  most recent CMB observations (see e.g. \cite{15}). Hence, the observational constraints obtained from CMB results may give us important information and constraints on the dynamics of the primordial string cosmology backgrounds. 

In this paper, following the basic postulate of ``Asymptotic Past Triviality" (APT) proposed to characterize the initial state of the pre-big bang scenario \cite{BDV}, we will consider a simple example of cosmic evolution with an initial regime described by a low-energy, homogeneous solution of the ten-dimensional  string effective action. Assuming that three dimensions are isotropically expanding we will show, in such a context, that the amplification of the quantum fluctuations of the axion background with a red spectral distribution is unavoidably associated with a phase of growing dilaton (hence, growing string coupling). We will then recall a basic result of standard (slow-roll) models of inflation, where the quantum backreaction of super-horizon perturbations with a red spectrum  usually implies a finite temporal extension towards the past of the inflationary regime. Finally, we will show that such a conclusion is evaded in the context of pre-big bang inflation where, in spite of the red spectrum, the backreaction of the axion perturbations (and of the quantum fluctuations of other possible background fields) is always negligible as we go back in time, without limits. 

We thus conclude that the cosmological perturbations imprinting the today   properties of the CMB radiation, at large-scales, may naturally emerge from a low energy and weak-coupling regime, with arbitrarily small initial dilaton, asymptotically approaching the string perturbative vacuum. Such a regime, on the other hand, may be typically an output -- through the mechanism of gravitational instability -- of a generic  (even inhomogeneous and spatially curved) initial state satisfying the most general APT assumptions \cite{BDV}.


\section{Red axion spectrum and growing string coupling}
\label{Sec2}
\setcounter{equation}{0}

In this paper we will consider the amplification of perturbations leaving the horizon during a higher-dimensional cosmological phase describing accelerated evolution and dynamical dimensional reduction. We shall work with a simple example of string-cosmology background with three isotropically expanding dimensions and six (internal) shrinking dimensions, not necessarily isotropic, described by the metric
\beq
ds^2= a(\eta)^2 \left(d\eta^2 - |d\vec x|^2\right) - \sum_i b_i^2(\eta) dy_i^2, ~~~~~~~~~ i=1, \dots, 6,
\label{1}
\eeq
where $\eta$ is the conformal time. In the low-energy regime, ranging from the initial time $\eta=-\eta_i$ to a limiting epoch $\eta =-\eta_s$, the background evolution can be appropriately described by the following dilaton-driven, vacuum solution of the tree-level string cosmology equations \cite{16},
\bea
&&
a(\eta) \sim (-\eta)^{\b_0\over 1- \b_0}, ~~~~~~~~~~~~~~~~~~~~~~~~~~~~~~ b_i(\eta) \sim  (-\eta)^{\b_i\over 1- \b_0},
\nonumber \\ &&
\phi(\eta) \sim { \sum_i \b_i +3 \b_0 -1\over 1-\b_0} \ln (-\eta),
 ~~~~~~~~~ -\eta_i \leq \eta \leq-\eta_s<0,
\label{2}
\eea
where the constant parameters $\b_0$, $\b_i$ satisfy the Kasner-like condition
\beq
3 \b_0^2+ \sum_i \b_i^2=1.
\label{3}
\eeq
Here $\phi$ is the ten-dimensional dilaton field, exponentially determining the string coupling which controls the topological loop expansion of the string action. We may regard the above model as the effective (string frame) description of a homogeneous and spatially flat patch of cosmic space-time emerging -- via gravitational instability -- from a generic ``past-trivial" initial vacuum state (see e.g. \cite{BDV}). 

At the final epoch $\eta=-\eta_s$ the string coupling parameter is still small, i.e $g^2= \exp(\phi) \ll1$, but the curvature reaches the string mass scale, $\Ms$, and the subsequent phase of background evolution is dominated by the contributions of higher-order $\ap$ corrections (see e.g. the solutions reported in \cite{17}). In this paper, however, we are only interested in perturbation modes of frequency low enough to be relevant at the large scales typical of the observed CMB anisotropy. Hence, for the purpose of this paper, it will be enough to consider the amplification of perturbations leaving the horizon at very early epochs, when the background evolution is still described by the low-energy solution (\ref{2}). 

During such a phase the axion background is trivial, $\sg=0$, but its quantum fluctuations $\da \sg$ are nonvaninshing, and their evolution is described by the (dimensionally reduced) effective action
\beq
S={1\over 2} \int d^4 x \sqrt{-g}\, \xi_\sg^2 \left( \da \sg^{\prime 2} +  \da \sg \nabla^2  \da \sg \right).
\label{4}
\eeq
(see e.g. \cite{2,18}). 
Here the prime denotes the derivative with respect to the conformal time $\eta$, and $\xi_\sg$ is the effective pump field defined by \cite{10,11}
\beq
\xi_\sg= a\left(\prod_i  b_i\right)^{-1/2} e^{\phi/2} \sim (-\eta)^{\a_\sg},  ~~~~~~~~
\a_\sg ={5 \b_0-1\over 2(1-\b_0)}
\label{5}
\eeq 
(we have used the solution (\ref{2}) to compute $\a_\sg$). Let us now introduce the canonical variable $u=\xi_\sg \da \sg$, which diagonalizes the kinetic part of the action (\ref{4}) and satisfies the Schrodinger-like equation
\beq
u^{\prime \prime}- \left( \nabla^2 +\xi_\sg^{-1} \xi_\sg^{\prime \prime} \right)u=0.
\label{6}
\eeq
The Fourier modes $u_k$, well outside the horizon ($k|\eta| \ll1$), are thus described by the asymptotic solution
\beq
u_k= A_k (-\eta)^{\a_\sg} + B_k (-\eta)^{1-\a_\sg}, ~~~~~~~~
|k\eta| \ll1, ~~~~~~~~ \eta<0
\label{7}
\eeq
(see e.g. \cite{19}), where $A_k$, $B_k$ are integration constants to be determined by normalizing the solutions inside the horizon (i.e for $|k\eta \gg1$) to the quantum fluctuations of the vacuum. One then finds, for such super-horizon modes, the general spectral distribution \cite{19}
\beq
\Da_\sg^2(k)= {k^3\over 2 \pi^2} \left|\da \sg_k\right|^2 \sim k^{n_s-1},
~~~~~~~~~~~~
n_s-1= 3- |2 \a_\sg -1|.
\label{8}
\eeq

Let us now discuss the constraints associated with a red axion spectrum, $n_s<1$, under the assumption that the three ``external" spatial dimensions are undergoing a phase of accelerated expansion, which imposes $\b_0/(1-\b_0)<0$. 

Let us first notice that the Kasner condition (\ref{3}) implies $\b_0^2 \leq 1/3$, namely
\beq
-{1\over \sqrt{3}} \leq \b_0 \leq {1\over \sqrt{3}},
\label{9}
\eeq
so that the expansion of the external space can only be obtained for negative values of $\b_0$ (the possibility $\b_0>1$ turns out to be excluded). Hence, the spectral index (\ref{8}) reduces to 
\beq
 n_s-1=2+2 \a_\sg= {3 \b_0+1\over 1-\b_0},
 \label{10}
 \eeq
and the condition of red spectrum ($n_s<1$), combined with Eq. (\ref{9}), leaves for $\b_0$ the allowed range
\beq
-{1\over \sqrt{3}} \leq \b_0 < -{1\over {3}}.
\label{11}
\eeq

The temporal evolution of the dilaton, on the other hand, is controlled by the combination of parameters $\sum_i \b_i+3\b_0-1$ (see Eq. (\ref{2})), and the allowed values of $\b_i$ are constrained by the Kasner condition (\ref{3}) which implies, using Eq. (\ref{11}), $\sum_i \b_i^2 \leq 2/3$. Assuming, for simplicity, that the extra dimensions are isotropic (with the same values of $\b_i$ along all internal directions), the above condition reduces to $\sum_i \b_i^2=6 \b_i^2 \leq 2/3$, and defines for $\b_i$ the allowed range
\beq
-{1\over {3}} \leq \b_i \leq {1\over {3}}.
\label{12}
\eeq

Combining the results (\ref{11}), (\ref{12}) we can now easily check that $1-3\b_0>2 \geq 6 \b_i$, so that 
\beq
{ \sum_i \b_i +3 \b_0 -1\over 1-\b_0}<0.
\label{13}
\eeq
It follows -- see Eq. (\ref{2}) -- that the class of backgrounds we are considering, which amplifies the quantum fluctuations of the axion  with a red spectrum,  unavoidably describes a phase where the dilaton $\phi$ (and the string coupling $g^2 = \exp \phi$) are monotonically growing in time. 
The currently observed value \cite{15} of the scalar spectral slope at large scales, $n_s \simeq 0.968$, is obtained, in particular, for 
\beq
\b_0 \simeq -0.348
\label{13a}
\eeq
(see Eq. (\ref{10})), and for arbitrary values of $\b_i$ lying in the range of Eq. (\ref{12}).


\section{Constraints on inflation due to the backreaction of 
super-horizon perturbations}
\label{Sec3}
\setcounter{equation}{0}

In view of the red spectral slope of the axion perturbations discussed in the previous section one might think that the regime of growing coupling, pre-big bang inflation, described by the string cosmology solution (\ref{2}), must be characterized by a finite duration and that, in particular, it cannot be extended back in time without limits. Indeed, this is exactly what happens in the context of  models of standard (slow-roll) inflation, where the backreaction of scalar super-horizon perturbations with a red spectrum grows as we go back in time, up to a limiting scale where the  energy density of the quantum perturbations becomes comparable with the energy density of the background and starts spoiling the inflationary dynamics.

It may be useful to briefly illustrate this effect by recalling, first of all, that in the standard inflationary scenario the amplification of a scalar metric perturbation $\psi$ is controlled by an effective pump field $\xi_\psi$ which evolves in time like the scale factor \cite{19},
\beq
\xi_\psi= { \dot \varphi\over H} a \sim a \sim (-\eta)^\a, ~~~~~~~~~~~~~ \eta<0
\label{14}
\eeq
($\varphi$ is the inflaton, $H= \dot a /a$ is the Hubble parameter, and the dot denotes differentiation with respect to the cosmic time $t$, defined by $dt=a d\eta$). During inflation the metric describes accelerated expansion, so that $\a<0$; the spectral distribution of the super-horizon scalar perturbations is thus given by $\Da_\psi^2 \sim k^{n_s-1}$, where $n_s-1=2+2\a$, and the slope is red ($n_s<1$) for $\a<-1$.

In order to evaluate the energy density of these  perturbations, outside the horizon and during inflation, it is convenient to introduce the canonical variable $u= \psi \xi_\psi$, and use the previous results (\ref{6}), (\ref{7}) for the Fourier mode $u_k$ (with the obvious replacements $\xi_\sg \ra \xi_\psi$ and $\a_\sg \ra \a$). 

For $\a<-1$ the asymptotic solution for $u_k$ is dominated by the first term of Eq. (\ref{7}) so that, outside the horizon, we have $u_k(\eta) \sim \xi_\psi(\eta)$, and $\psi_k \sim $ const. By imposing on $u_k$ the sub-horizon normalization to an initial spectrum of quantum fluctuations of the vacuum we can then write the asymptotic solution, modulo a phase, as
\beq
u_k \simeq {1\over \sqrt{2 k}} (-k\eta)^\a, ~~~~~~~~~~
k |\eta| \ll 1, ~~~~~~~~~~ \eta<0.
\label{15}
\eeq
We can then apply the results for the vacuum expectation value of the energy density of the super-horizon modes reported in \cite{20}, assuming that the inflationary regime starts at an initial epoch $-\eta_i <0$, considering a generic epoch $\eta$ with $-\eta_i <\eta<0$, and integrating over all super-horizon modes from $k_i=\eta_i^{-1}$ to $k=-\eta^{-1}>k_i$. Taking into account that $\psi_k \sim $ const for $k|\eta|\ll1$, we then obtain that the leading contribution to the energy density $\r_\psi(\eta)$ of the amplified scalar perturbations, outside the horizon and during inflation, is given by \cite{20}
\beq
\r_\psi(\eta) \sim{1\over a^4} \int_{\eta_i^{-1}}^{-\eta^{-1}} dk k^4 |u_k|^2
= H^4 \left[c_1 +c_2 \left(-\eta\over \eta_i\right)^{4+2\a}\right],
\label{16}
\eeq
where $c_1$ and $c_2$ are dimensionless integration constants of order one (we have used Eq. (\ref{15}) for $u_k$, and the relation $H \sim (-a\eta)^{-1}$ for the Hubble parameter).

Let us now consider an inflationary phase producing a slightly red spectrum, with $|n_s-1| \ll 1$. We can put, in this case, $\a=-1-\ep$, where $\ep$ is a small and constant (slow-roll) parameter such that $0< \ep \ll1$. It turns out that $4+2\a>0$ so that, during inflation (i.e. for $\eta >- \eta_i$), the time evolution of the perturbation energy $\r_\psi(\eta)$ is well approximated by the relation $\r_\psi \simeq H^4$. The corresponding Hubble parameter, on the other hand, is slightly decreasing for $\eta \ra 0_-$, since
\beq
H \sim (-a\eta)^{-1} \sim (-\eta)^{-(1+\a)} \sim (-\eta)^\ep , 
~~~~~~~~~~~~~~~~~~~ \eta <0.
\label{17}
\eeq
It follows that $\r_\psi$ tends to grow as we go back in time, and that it becomes of the same order as the background energy density when $\r_\psi \simeq H^4 \simeq H^2 \Mp^2$, where $\Mp$ is the Planck mass. After that time the model of background evolution is completely spoiled by the large backreaction of the super horizon perturbations: as a consequence, the associated inflationary regime has a limited past extension.

The maximal allowed duration of the inflationary regime, in this context, can be 
easily estimated by computing  the corresponding number $N$ of e-foldings, at any given epoch $\eta>-\eta_i$, by assuming that inflation starts at the initial epoch $\eta_i$ of maximal allowed backreaction  (such that $H_i^4  \simeq H_i^2 \Mp^2$). Let us assume, for the sake of simplicity, that the given model of slow-roll inflation can be extrapolated without any (higher-curvature, quantum) correction\footnote{If we include such corrections (see e.g. \cite{q1,q2}) the allowed number $N$ of e-foldings given in Eq. (\ref{19}) may change, but the limiting effects on the duration of inflation are qualitatively the same.} up to the limiting scale $H_i =\Mp$. We thus immediately obtain 
\beq
e^N={a\over a_i}=\left(-\eta\over \eta_i\right)^{-(1+\ep)}\laq \left(H\over \Mp\right)^{-{1+\ep\over \ep}},
\label{18}
\eeq
which implies
\beq
N \laq {1\over \ep}\ln \left(\Mp\over H\right), ~~~~~~~~~~~~~~~
\ep  \ll 1.
\label{19}
\eeq

As we shall see in the following section, the above temporal limitation towards the past of the inflationary regime does not apply to the phase of pre-big bang inflation illustrated in Sect. \ref{Sec2}, in spite of the amplification of the axion fluctuations with a red spectrum. Their backreaction, in fact, turns out to be always negligible since the  evolution in time of the axion pump field is different from that of the scale factor, and is thus differently related to the evolution of the Hubble parameter.

\section{Backreaction of perturbations in pre-big bang inflation}
\label{Sec4}
\setcounter{equation}{0}

Let us now apply the procedure of the previous section in order to compute the energy density of the super-horizon axion fluctuations, amplified by the phase of pre-big bang inflation described in Sect. \ref{Sec2}.

Given the allowed values of $\b_0$ (see Sect. \ref{Sec2}), the power of the axion pump field (\ref{5}) is negative, $\a_\sg<0$, and the asymptotic solution for the canonical variable $u_k$ is dominated by the first term of Eq. (\ref{7}). The normalized solution for $u_k$ takes then the form (\ref{15}) (with the obvious replacement $\a \ra \a_\sg$). For such modes we have, outside the horizon, $u_k(\eta) \sim \xi_\sg(\eta)$ and $\da \sg \sim $ const, so that the leading contribution to the energy density $\r_\sg$ of the axion fluctuations, outside the horizon and during the phase of pre-big bang inflation, can be expressed exactly as in Eq. (\ref{16}) (again, with the replacement $\a \ra \a_\sg$). Finally, given the value of $\b_0$ reproducing the observed (slightly red) slope of scalar perturbations (see Eq. (\ref{13a})), we have
\beq
4+2 \a_\sg ={3+\b_0\over 1-\b_0}>0.
\label{20}
\eeq
Hence, at any given epoch $\eta$ during the phase of pre-big bang inflation (namely, at any time $\eta$ such that $-\eta_i<\eta<0$), the second term of Eq. (\ref{16}) is negligible, and the time evolution of the axion energy density is well approximated by
\beq
\r_\sg(\eta) \simeq H^4, ~~~~~~~~~~~~~~
-\eta_i \laq \eta <0,
\label{21}
\eeq
exactly as before for $\r_\psi(\eta)$. 

The contribution of $\r_\sg$ to the background geometry, however, is now controlled by the (dimensionally reduced) string cosmology equations, which imply \cite{2,18}
\beq
H^2 \simeq G_4(\eta) \r_\sg(\eta) = {e^{\phi_4(\eta)}\over \Ms^2} \r_\sg(\eta).
\label{22}
\eeq
Here $G_4$ is the effective gravitational coupling in four dimensions, determined by the string mass $\Ms$ and by the effective four-dimensional string coupling parameter $g_4$, defined by
\beq
g_4^2(\eta) \equiv e^{\phi_4} = \left(\prod_i b_i\right)^{-1} e^\phi \sim (-\eta) ^{3 \b_0-1\over 1-\b_0}
\label{23}
\eeq
(we have used the explicit form of the Kasner solution (\ref{2})). The relative importance of the axion backreaction is thus controlled by the ratio $G_4 \r_\sg/H^2$ which evolves in time as
\beq
{g_4^2 \r_\sg\over H^2 \Ms^2} \sim (-\eta)^{-3}, ~~~~~~~~~~~~
-\eta_i \leq \eta<0,
\label{24}
\eeq
and which {\em decreases} as we go back in time, differently from the backreaction of the scalar perturbations in the standard inflationary scenario of Sect. \ref{Sec3}.

At the end of the low-energy phase described by the Kasner solution (\ref{2}), i.e. at the epoch $\eta=-\eta_s$ where the background reaches the string scale, $H(\eta_s)\equiv H_s= \Ms$, the backreaction of the axion fluctuations reaches the highest  intensity $(g_4^2 H^2/\Ms^2)_{\eta_s}= g_4^2(\eta_s) \equiv g_s^2$, but it is still small since, at that epoch, the string coupling is still perturbative, $g_s^2 \ll1$ (see e.g. \cite{17}). At all previous epochs $\eta <-\eta_s$ the axion contributions are even smaller,
\beq
{g_4^2 H^2\over\Ms^2}= g_s^2 \left(-\eta\over \eta_s\right)^{-3} \laq g_s^2 \ll1, ~~~~~~~~~~~~~ -\eta_i \leq \eta \leq -\eta_s,
\label{25}
\eeq
and always negligible. Hence, the phase of pre-big bang inflation described by Eq. (\ref{2}) can be extended back in time without limits ($-\eta_i \ra -\infty$), in the absence of any past-temporal boundary imposed by the presence of super-horizon axion fluctuations amplified with a red spectral distribution (differently from the case of scalar perturbations in the standard inflationary context).

The same conclusion applies if we consider the backreaction of scalar and tensor metric perturbations directly amplified during the phase of pre-big bang inflation. They are characterized, in fact, by a steep blue spectrum which is a consequence of the growth of the curvature scale -- like the blue spectra produced in the standard inflationary context. As a consequence, the contributions of their energy density automatically decrease as we go back in time, and goes to zero for $\eta \ra -\infty$. 

This is the case, also, if we consider the amplified perturbations of the internal and ``mixed" components of the higher-dimensional metric, as well as the extra-dimensional components of the NS-NS two form whose four-dimensional sector corresponds to the Kalb-Ramond axion. For all these different fluctuations the associated pump fields have different time dependence, and it may even happen that the asymptotic solutions for the canonical variables are dominated by the second term of Eq. (\ref{7}) (corresponding to fluctuations which are growing outside the horizon). However, given the constraints on $\b_0$, $\b_i$ required to obtain a viable axion spectrum (see Sect. \ref{Sec2}), we have checked that, in all cases,  the spectra of these extra-dimensional fields turns out to be blue, that their energy density evolves in time like $H^4$, and that their backreaction goes to zero as we go back in time, without imposing any temporal limit on the extension towards the past of the considered phase of pre-big bang inflation.

\section{Conclusion}
\label{Sec5}
\setcounter{equation}{0}

In the context of pre-big bang cosmology, the production of a primordial spectrum of scalar perturbations with a slightly red slope, in agreement with present observations, may be associated with an initial low-energy phase where the dilaton and the string coupling are monotonically growing. 

In spite of their red spectrum, the quantum backreaction of the super-horizon modes does not affect at all the given background dynamics, so that the phase of inflationary pre-big bang evolution can be extended back in time, towards smaller and smaller values of the string coupling. A realistic inflationary model, in such a context, may thus  emerge as the result of classical gravitational instability from a  generic, past-trivial, string vacuum state \cite{BDV}. The ``naturalness" of such a scenario will be discussed (following the approach presented in \cite{BDV}) in a forthcoming paper \cite{25}.

 
\section*{Acknowledgements}

This work is supported in part by INFN under the program TAsP (Theoretical Astroparticle Physics).  It is a pleasure to thank Gabriele Veneziano for many useful discussions. 
I wish to thank, also, an anonymous Referee of JCAP for his/hers comments concerning a recently published paper \cite{23}, and for his/hers constructive criticism  which has motivated, in part, the study of super-horizon perturbations and of their backreaction presented in this paper.


\end{document}